\documentclass[final,3p,times,twocolumn]{elsarticle}

\usepackage{amssymb}
\usepackage{amsmath}
\usepackage{rotating}
\usepackage{dirtytalk}
\usepackage{caption}
\usepackage[dvipsnames]{xcolor}

\journal{Nuclear Instrumentation and Methods: A}

\begin{document}

\begin{frontmatter}



\title{Measurement of Charge State Distributions using a Scintillation Screen}


\author[ou,frib]{C. Marshall \fnref{fn1}}
\author[ou]{Z. Meisel \fnref{fn2}}
\author[frib]{F. Montes}
\author[frib]{L. Wagner \fnref{fn3}}
\author[frib]{K. Hermansen}
\author[frib]{R. Garg}
\author[ornl]{K.A. Chipps}
\author[cmu]{P. Tsintari}
\author[cmu]{N. Dimitrakopoulos}
\author[nd]{G.P.A. Berg}
\author[ou]{C. Brune}
\author[nd]{M. Couder}
\author[csm]{U. Greife}
\author[frib]{H. Schatz}
\author[ornl]{M. S. Smith}

\fntext[fn1]{Present Address: Department of Physics and Astronomy, University of North Carolina at Chapel Hill, Chapel Hill, NC, USA}
\fntext[fn2]{Present Address: Battelle, Columbus, OH, USA}
\fntext[fn3]{Present Address: TRIUMF, Vancouver, British Columbia, Canada}

\affiliation[ou]{organization={Institute For Nuclear \& Particle Physics, Ohio University},
            city={Athens},
            state={OH},
            country={USA}}

\affiliation[frib]{organization={Facility For Rare Isotope Beams},
            city={East Lansing},
            state={MI},
            country={USA}}

\affiliation[ornl]{organization={Physics Division, Oak Ridge National Laboratory},
            city={Oak Ridge},
            state={TN},
            country={USA}}

\affiliation[cmu]{organization={Department of Physics, Central Michigan University},
            city={Mt. Pleasant},
            state={MI},
            country={USA}}

\affiliation[nd]{organization={Department of Physics and Astronomy, University of Notre Dame},
            city={Notre Dame},
            state={IN},
            country={USA}}

\affiliation[csm]{organization={Department of Physics, Colorado School of Mines},
            city={Golden},
            state={CO},
            country={USA}}


\begin{abstract}
Absolute cross sections measured using electromagnetic devices to separate and detect heavy recoiling ions need to be corrected for charge state fractions. Accurate prediction of charge state distributions using theoretical models is not always a possibility, especially in energy and mass regions where data is sparse. As such, it is often necessary to measure charge state fractions directly. In this paper we present a novel method of using a scintillation screen along with a CMOS camera to image the charge dispersed beam after a set of magnetic dipoles. A measurement of the charge state distribution for $^{88}$Sr passing through a natural carbon foil is performed. Using a Bayesian model to extract statistically meaningful uncertainties from these images, we find agreement between the new method and a more traditional method using Faraday cups. Future work is need to better understand systematic uncertainties. Our technique offers a viable method to measure charge state distributions.  

\end{abstract}




\end{frontmatter}


\section{Introduction}
\label{sec:intro}

The direct measurement of nuclear reaction cross sections for short-lived isotopes presents several challenges. Beam intensities are orders of magnitude lower than stable beam experiments, reactions must be performed in inverse kinematics, and beam induced backgrounds can inhibit the detection of prompt decays from the reaction of interest. In order for measurements to be sensitive to lower cross sections they must combine high detection efficiencies and background suppression, all in a setup suitable for inverse kinematics studies. 

Efficiency can be gained by detecting the heavy recoil from the reaction; however, inverse kinematics imply extremely forward-focused recoils that spatially overlap with the beam, swamping most detection methods one might consider for the recoils. Down-scattering beam creates overlaps in both energy and momentum, limiting the suppression offered by systems only employing momentum analysis \cite{Ruiz_2014}. Recoil separators address these challenges by combining momentum and velocity analysis, achieved in practice by combining magnetic dipoles with: electric dipoles \cite{CORMIER_1981, DAVIDS_1989, SPOLAORE_1985, DAVIDS_2005, COLE_1992, MORINOBU_1992, HUTCHEON_2003}, velocity filters \cite{HARWOOD_1981, Hahn_1987, JAMES_1988, TRIBBLE_1989, COUDER_2003}, and/or measurement of the recoil time-of-flight (TOF) \cite{WOLLNIK_1987}.

Nuclei emerging from a target will have a mixture of charge states due to the probabilistic loss and capture of electrons according to the charge changing cross sections. The mixture is known as a charge state distribution (CSD).
Since recoil separators are electromagnetic devices, they can only transmit charge states of the recoil that fall within the magnetic rigidity ($B \rho$) and electric rigidity ($E \rho$) limits of the system. If only one charge state is transmitted, as is the case for separators designed to provide the highest levels of beam rejection, it is key to know the precise fraction of the outgoing recoils that have the selected charge state in order to calculate the cross section. 

While several theoretical \cite{ROZET_1996, lamour_2015, WINCKLER_2017} and phenomenological codes \cite{sayer_1977, baudinet_1982, SHIMA_1992, LIU_2003} exist that can calculate CSDs, reliance on theoretical models introduces the potential for uncontrolled systematic errors unsuitable for absolute cross section measurements. This situation is especially concerning at the low energies important to nuclear astrophysics where data on CSDs is lacking \cite{LIU_2003}. As a result, direct measurements of charge state distributions remain an essential ingredient for accurate cross section measurements using recoil separators.

In this paper we present a new method for measuring charge state distributions using scintillation screens. The technique is demonstrated by measuring CSDs for strontium ($^{88}$Sr) through carbon at beam energies of $2.5$, $2.75$, and $3.0$ MeV/u. We use the SEparator for CApture Reactions (SECAR) recoil separator \cite{BERG_2018} located within ReA3 \cite{LEITNER_2013}, the low energy reaccelerated beam portion of the Facility for Rare Isotope Beams (FRIB). SECAR is designed to study $(p, \gamma)$ reactions up to $A = 65$ using the JENSA gas-jet target system \cite{Chipps_2014, BARDAYAN_2015, SCHMIDT_2018}, but its acceptance and maximum rigidities allow measurements on much heavier nuclei provided that the desired recoils have a larger mass difference from the beam. Higher mass reactions have larger on average $B \rho$ values that push the limits of the system's design. Their reaction cross sections tend to vary slowly as a function of energy causing recoils to be produced throughout the length of the gas target. Under such conditions, gas targets have two distinct drawbacks: they produce an average charge state that is lower relative to solid targets \cite{sayer_1977, VOCKENHUBER_2007}, which can cause the most highly populated charge states to be outside of the rigidity limit of SECAR, and a CSD that can depend on where a recoil is produced along the target length. A post-target carbon foil remedies both of these issues by raising the average charge state of the recoils and ensuring they all pass through a sufficient thickness of material to establish charge state equilibrium \cite{SCHURMANN_2004, VOCKENHUBER_2007}. 

The prominent role CSDs play in cross section measurements using recoil separators along with the potential to push SECAR beyond its designed mass range leads to the need for complementary methods to extract CSDs for a wide variety of ions. To this end, we present a novel method for measuring CSDs using a scintillation screen. A Bayesian model is developed and applied to the data in order to extract reliable statistical uncertainties. A more in depth study of systematic uncertainties is needed in the future. The present study finds measurements with Faraday cups and scintillation screens to be in good agreement.

\section{Experimental Method}
\label{sec:exp-method}

\begin{figure*}
    \centering
    \includegraphics[width=0.75\textwidth]{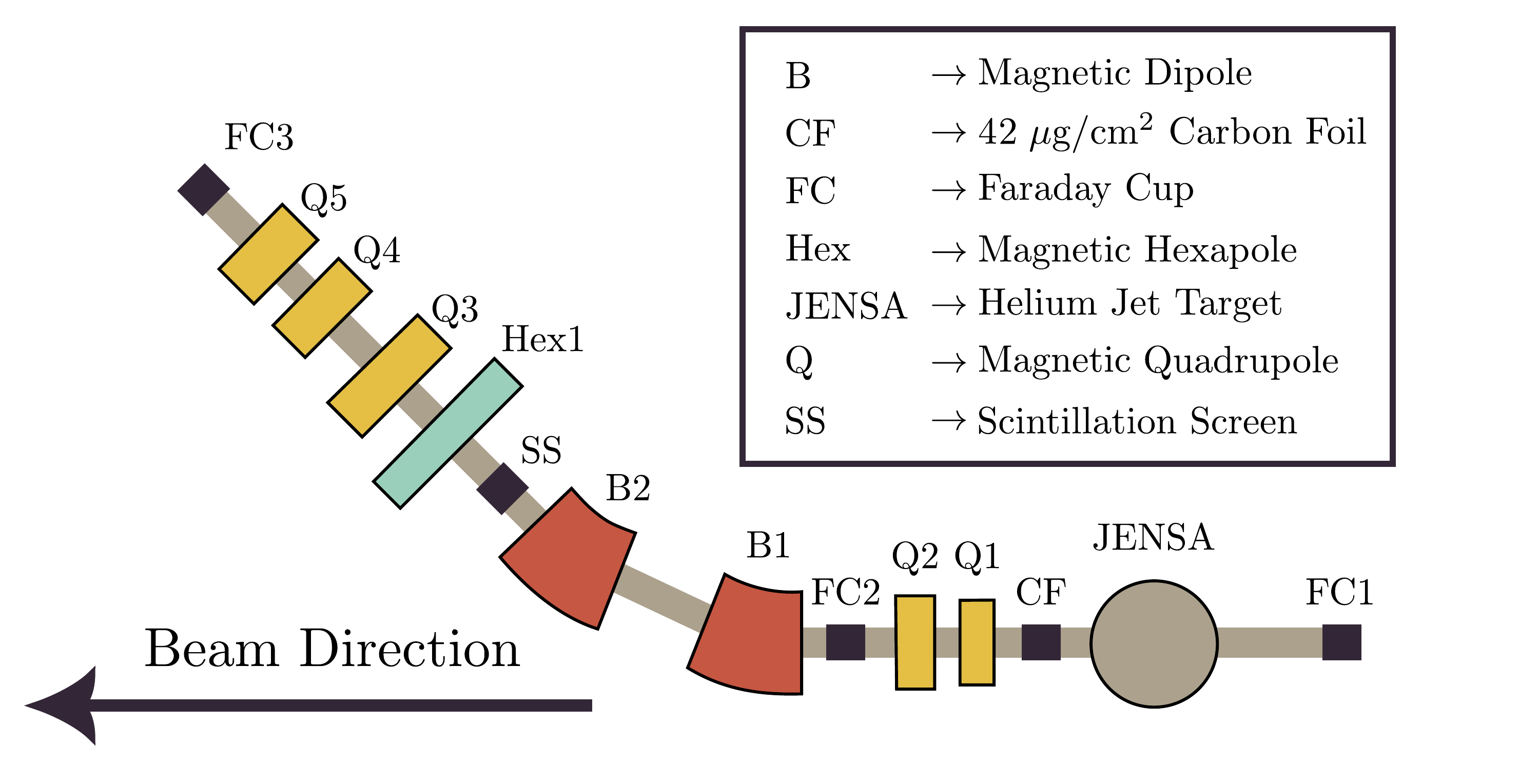}
    \caption{Sketch of SECAR up to the first focal plane which is located at FC3. Elements used for the charge state distribution measurements are represented by dark squares. All acronyms are defined in the figure.}
    \label{fig:secar_drawing_fp1}
\end{figure*}

ReA3 delivered a beam of $^{88}$Sr$^{28+}$ at energies of $1.75$, $2.25$, and $2.75$ MeV/u. Beam intensities were typically on the order of $1-4$ epA ($\approx 10^5$ pps). A $42(4)$ $\mu$g/cm$^2$ carbon foil purchased from the Arizona Carbon Foil Company \cite{acfmetals} was used as the post target stripper foil. The foil was mounted on a linear motion feedthrough downstream of the gas-jet target. SECAR is designed to accept recoiling ions from a $1.5$ mm diameter beam spot with angles up to $\pm 25$ mrad. In order for the stripper foil to be compatible with these requirements, it is necessary for it to cover a diameter of $\approx 2.5$ cm. As such our foil was mounted on a frame with a diameter of $2.7$ cm. Mounting $\approx \! 40$ $\mu$g/cm$^2$ foils of this diameter was found to be challenging. To increase the success rate of floating and mounting the foils on the frame, foils with a Collidon coating were used. Ref.~\cite{LIECHTENSTEIN_2006} found exposing a Collidon coated carbon foil to a $^{16}$O$^{2+}$ beam with energy $6$ MeV, $50 \text{-} 100$ enA intensity, and beam spot size of $16$-mm$^2$ removed the coating instantaneously. The three orders of magnitude difference in beam current between that experiment and ours is not compensated for by the difference in stopping power or beam spot size. With this disparity, the Collidon coating could be expected to survive several minutes of beam exposure. 

 The so-called charge selection stage of SECAR used for this experiment is sketched in Fig.~\ref{fig:secar_drawing_fp1}. A single charge state of the desired mass is selected at an energy dispersed focus occurring at the location of the third Faraday cup, FC3, which is after the the first set of magnetic dipoles (B1/B2). The fields of B1/B2 are set, matched, and monitored via two separate NMR probes. The dipoles had previously been energy calibrated using the $992$-keV resonance of $^{27}$Al$(p, \gamma)$. Measuring the current prior to JENSA with FC1 and after B1/B2 with FC3  makes it possible to perform an absolute measurement of the charge state fraction. Measuring charge state fractions by utilizing the charge selection stage of a recoil separator is a well proven method \cite{LIU_2003}. Currents read on these cups can be related to the charge state fraction, $F_q$, of the charge state selected at FC3 ($q_{\text{FC3}}$) via: 
\begin{equation}
    \label{eq:fc_charge_state}
    F_q = \frac{q_{\text{FC1}}}{q_{\text{FC3}}} \frac{I_{\text{FC3}}}{I_{\text{FC1}}}, 
\end{equation}
where $q$ refers to the charge state of the beam ($q_{\text{FC1}}$) or the charge state selected using B1/B2 ($q_{\text{FC3}}$) and $I$ denotes the current measured at either FC1 or FC3. The sum of $F_q$, by definition, is $1$:

 \begin{equation}
     \label{eq:csd_sum}
     \sum_q F_q = 1.
 \end{equation}

As can be seen in Fig.~\ref{fig:secar_drawing_fp1}, FC1 measures current prior to transmission through the JENSA chamber. Transmission losses between this Faraday cup and the one after the chamber, FC2, can lead to an incorrect measurement of $F_q$. Transmission losses are frequently a result of the beam striking the gas restricting apertures used for differential pumping located both upstream and downstream of the central chamber. By measuring the ratio of currents on FC1 and FC2 with an empty target and retracted stripper foil, a transmission efficiency can be deduced:
\begin{equation}
    \label{eq:fc_eff}
    \eta_{\text{Beam}} = \frac{I_{\text{FC2}}}{I_{\text{FC1}}}, 
\end{equation}
and in turn Eq.~\ref{eq:fc_charge_state} can be corrected for transmission losses. However, this simple procedure is inappropriate for losses downstream of JENSA since any additional source of angular scattering (i.e. straggling through the target or carbon foil) will lead to additional losses not accounted for by Eq.~\ref{eq:fc_eff}. 

\begin{figure*}
    \centering
    \includegraphics[width=0.7\textwidth]{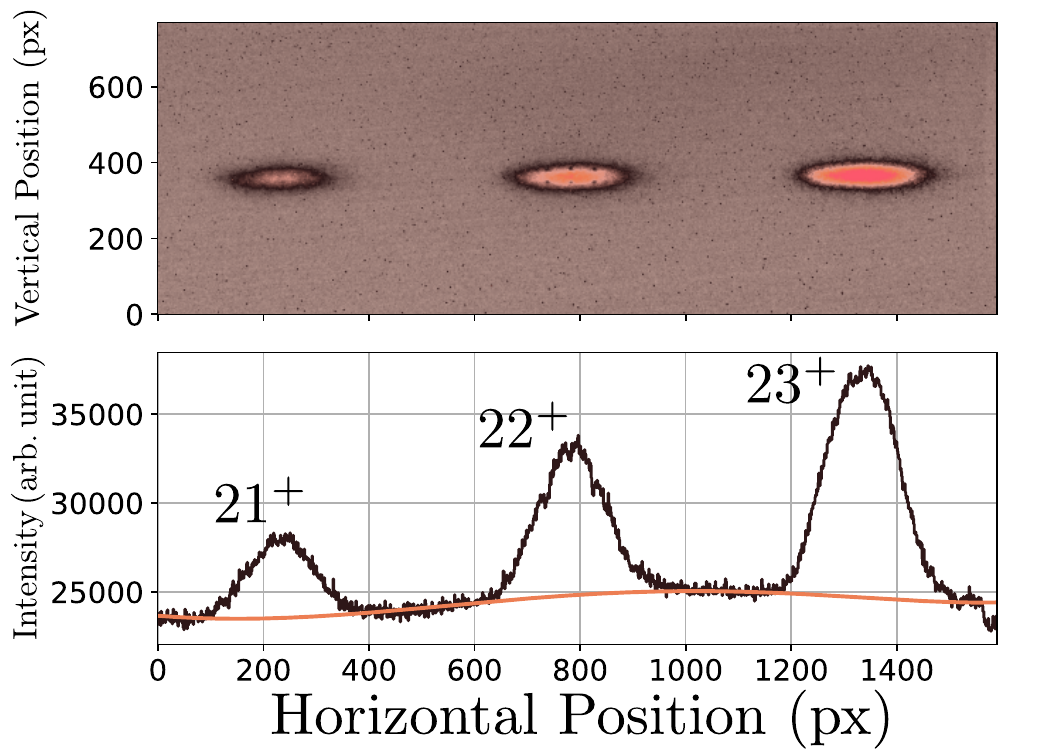}
    \caption{Image obtained from a CMOS camera imaging the scintillation screen 1 m downsteam of the first set of magnetic dipoles of SECAR. The beam has passed through a thin carbon foil. The $^{88}$Sr beam energy is $1.75$ MeV/u and the observed charge states from image/beam left to right are $21^{+}$, $22^{+}$, and $23^{+}$. An example of the background subtraction discussed in Sec.~\ref{sec:data-reduction} is shown as a solid line in the bottom panel.}
    \label{fig:example_image}
\end{figure*}

Our novel scheme to measure a CSD utilizes the same section of SECAR as described above. Looking again at Fig.~\ref{fig:secar_drawing_fp1}, we can see that FC3 is not immediately after B2, but is instead after 4 additional magnetic elements.
However, about 1 m after B2, but before the first hexapole (Hex1), the dispersion of the charge states is roughly half of that at the focal plane \cite{BERG_2018}. A scintillation screen is located at this position and is able to simultaneously view three charge states for $\Delta q / q < 5 \%$. By stepping the fields of B1/B2 to center successive charge states, we create a series of overlapping images that span over the detectable charge state fractions. The scintillation screen is a copper plate coated with P22-R (Y$_2$O$_2$S:Eu$^{3+}$). A CMOS camera, DMK 33GX174 from Imaging Source \cite{camera}, records the scintillation of the screen when beam impinges on it. An example image is shown in Fig.~\ref{fig:example_image}. 

 It is clear that qualitative information about the charge state distribution is present in Fig.~\ref{fig:example_image}, but a quantitative analysis requires a statistical model to relate multiple images, with each image subject to unobserved fluctuations in beam intensity.

\section{Analysis Methods}
\label{sec:bay-model}
 Before discussing our data reduction and statistical model, it is necessary to reference the assumptions that have been made in our analysis. They are:

\begin{enumerate}
    \item The light output varies linearly with the incident beam intensity.
    \item  The CMOS sensor produces images with intensity linearly related to the light output of the viewer.
    \item The shape of the background does not vary between images. 
\end{enumerate}

Issues with items $1$ and $2$ will be explored in Sec.~\ref{sec:linearity} for the measurements at $1.75$ MeV/u when the CMOS was found to saturate. Background images taken over the course of the measurement did not show any indication that item $3$ is  invalid.

\subsection{Data Reduction}
\label{sec:data-reduction}

Images, like those of Fig~\ref{fig:example_image}, were taken for each $B1/B2$ setting that centered a charge state on the screen. The gray-scale pixel values, which correspond to photon intensity, were projected on to the axis that corresponded to the dispersive plane of the dipoles. Quantitative analysis was then performed on the three separate peak regions. Background subtraction for these peaks was done by using images taken without beam striking the scintillation screen. These background images were found to be well described by a $5^{\textnormal{th}}$ degree polynomial. The non-intercept terms of this polynomial, i.e. those terms of $x^{n}$ with $n > 0$, were fit to the background images and then held fixed for the image analysis of the charge states. The intercept was allowed to vary for each individual image and was estimated from the regions between the peaks. 

After background subtraction the peak areas were found from summing the counts in a specified region. Statistical error was negligible, with the weakest peaks still containing $> 10^4$ counts. Since the peak regions are defined by \say{eye}, larger variation is found from the choice of the region to integrate. Nevertheless, investigations of this effect found fluctuations on the deduced number of counts to be around $3 \%$, which also provides supporting evidence to the efficacy of our background subtraction procedure.  

\subsection{Bayesian Normalization Method}
\label{sec:bay-norm-model}

Our challenge is to extract charge state fractions with meaningful uncertainties given beam fluctuations, the arbitrary normalization of the peak intensities, and possible issues with the CMOS response. For this work, we approach these challenges using a Bayesian model, which by construction will provide statistically meaningful uncertainties. A Bayesian model can be specified by assigning probability distributions to each model parameter and by using Bayes' theorem, which is given by: 
\begin{equation}
  \label{eq:bayes_theorem}
  P(\boldsymbol{\theta}|\mathbf{D}) = \frac{P(\mathbf{D}|\boldsymbol{\theta}) P(\boldsymbol{\theta})}
  {P(\mathbf{D})},
\end{equation}
where $P(\boldsymbol{\theta})$ are the prior probability distributions of the model parameters $\boldsymbol{\theta}$, $P(\mathbf{D}|\boldsymbol{\theta})$ is the likelihood function for data $\mathbf{D}$, $P(\mathbf{D})$ is the evidence, and $P(\boldsymbol{\theta}|\mathbf{D})$ is the posterior \cite{bayes}. Briefly, prior probabilities are assigned based on beliefs about the model parameters before the data are considered, the likelihood expresses the probability of observing the data given a set of model parameters, and the evidence is an overall normalization. For the purpose of extracting the posterior distributions of charge state fractions, only the priors and likelihood are necessary while the evidence amounts to a normalization constant that can remain unknown. We now turn our attention to defining the necessary prior and likelihood distributions.

In the absence of accurate knowledge of the response of the CMOS and scintillation screen, only relative intensities can be established. Additionally, the CMOS and scintillation screen will be unable to detect beam intensities below a certain threshold. If charge states from $q_{min}$ to $q_{max}$ are observed and there is sufficient evidence to assert that $\sum_{q_{min}}^{q_{max}} F_q \approx 1$, then a normalization procedure can be carried out to extract CSFs:
\begin{equation}
    \label{eq:two_step_sum}
    F_q = \frac{I_q}{\sum_{q_{min}}^{q_{max}} I_i},
\end{equation}
where $I_q$ is the measured light intensity of the charge state $q$, the index $i$ runs from the lowest observed charge state, $q_{min}$, to the highest charge state, $q_{max}$.

If every charge state was measured in a single image, Eq.~\ref{eq:two_step_sum} could be used to directly estimate the relative charge state fractions. In practice, each image contains at most three charge states (Sec.~\ref{sec:exp-method}), and the normalization cannot be assumed to be constant due to intensity fluctuations of the beam. Intermediate normalization of each image is therefore required before the application of Eq.~\ref{eq:two_step_sum}. We denote each image using the index $k$. By stepping the charge state states one at a time, such that an image is guaranteed to share at least two charge states with the previous image, an image averaged intensity, $\Bar{I}_q$, can be found:
\begin{equation}
    \label{eq:average_intensity}
    \phi_k I_{\{k, q\}} \sim \mathcal{N}( \Bar{I}_q, \sigma_q^2),
\end{equation}
where $\sim$ means \say{distributed according to}, $I_{\{k, q\}}$ is the measured intensity for charge state $q$ for the image $k$, $\phi_k$ is the normalization factor for the $k^{\textnormal{th}}$ image. The product $\phi_{k} I_{\{k, q\}}$  is assumed to be normally distributed with a variance $\sigma_q^2$ and mean $\Bar{I}_q$. Assuming Poisson noise on the peak areas for each image implies image specific standard deviations $\ll 1 \%$, leading to the further assumption that the dominant source of uncertainty is a global fractional uncertainty, $\lambda$. An informative half normal prior is assigned to this parameter:
\begin{equation}
    \label{eq:lambda_prior}
    \lambda \sim \textnormal{HalfNorm}(0.1^2),
\end{equation}
where $\textnormal{HalfNorm}(0.1^2)$ is the half normal distribution with $\sigma = 0.1$, or, in other words, we expect the normalized intensities to vary by roughly $10 \%$ although larger variations have a non-zero probability. With this choice $\sigma_q$ becomes $\sigma_q = \lambda \Bar{I}_q$. 

Before the Bayesian model is complete a prior must be assigned to each $\phi_k$. At this point it must be noted that our normalization is completely arbitrary, meaning that our problem has infinitely many equivalent solutions. A somewhat simple choice is to scale the intensities relative to the most intense observed peak, and fix the normalization of the corresponding image, $k_{max}$. Each image can still take on any value (i.e. the most intense peak need not be the most intense after the other images have been normalized). As such, $\phi_k$ can be assigned a uniform prior over the interval $[0, 1/I_{min})$ so that the minimum observed intensity is allowed the freedom of becoming the most intense if that is implied by the data. 

Our full model is:
\begin{align}
  \label{eq:csd_model_two_step}
 & \textnormal{Priors:} \nonumber \\
 & \Bar{I}_{q} \sim \textnormal{Uniform}(0, 1/I_{min}) \nonumber \\
 & \phi_{k} \sim \textnormal{Uniform}(0, 1/I_{min}) \nonumber \\
 & \lambda \sim \textnormal{HalfNorm} (0.1^2) \nonumber \\
 & \textnormal{Functions:}  \\
 & I_{q, k}^{\prime} = \phi_k I_{q, k}^{exp}   \nonumber \\
 & \textnormal{Likelihood:} \nonumber \\
 & I_{q, k}^{\prime} \sim \mathcal{N}(\Bar{I}_q, \{ \lambda \Bar{I}_q \}^2)  \nonumber
 \end{align}

Application of the above gives a set of $\Bar{I}_q$ that can be renormalized using Eq.~\ref{eq:two_step_sum}. We also investigated an alternative method that implicitly uses the constraint $\sum_q F_q = 1$ to extract charge state fractions without the renormalization step. Details can be found in Appendix A. It was found to be in agreement with the above method, and it was not explored further in this work.  

\subsection{Faraday Cup Analysis}
\label{sec:fc_analysis}

In addition to the viewer images, Faraday cup measurements were made at each energy using FC1 and FC3 for purposes of comparison to the scintillation screen. The dimensions of the cups are designed to provide accurate charge integration to $1 \%$. At the low currents of this experiment ($0.5-10$ pA) the performance of the ammeters was also considered. A Keithley Model 6514 picoammeter was used during the experiment \cite{ammeter}. After the experiment the response of the ammeter was tested with a Keithley 261 picoampere source \cite{current-source}. Uncertainties were estimated based on the fluctuations observed in the ammeter readings as well as the offsets, and found to be below $1 \%$. Uncertainties are conservative in this case because the measurements are a convolution of effects from both the ammeter and current source.  

Due to the low uncertainties expected from loss of electrons and current integration, our uncertainties will be dominated by beam fluctuations and the number of readings used to estimate the current. Quantitative knowledge of these effects is difficult, if not impossible, to estimate; thus, we adopt a conservative $7 \%$ uncertainty based on a review of several hundred cup readings taken in subsequent experiments.   

Absolute measurements are possible using Faraday cups, but were avoided in this case due to the low beam currents, potential transmission losses after the carbon foil, and long $RC$-type discharge observed on FC3 making the offset of the cup difficult to obtain. It can be seen that normalizing the charge state fractions according to Eq.~\ref{eq:two_step_sum} will tend to diminish the effects of systematic errors that scale as a percentage. Uncertainties were propagated through Eq.~\ref{eq:two_step_sum} by assuming normal distributions for the cup readings, with standard deviations of $7 \%$ following the estimate given above, and using a Monte-Carlo method. $1 \times 10^6$ samples were drawn for each cup reading, currents were corrected for the charge state, and the final charge state fractions were calculated from the $68 \%$ credibility intervals for each set of samples renormalized using Eq.~\ref{eq:two_step_sum}.   

\section{Results}

The posterior distributions for the Bayesian model of Sec.~\ref{sec:bay-model} were estimated using Markov Chain Monte Carlo (MCMC). MCMC sampling was carried out using affine invariant slice sampling as implemented in the \texttt{Python} package   
\texttt{zeus} \cite{karamanis2020ensemble, karamanis2021zeus}.

Faraday cups readings, Sec.~\ref{sec:fc_analysis}, were used as a cross check of the charge state fractions and were carried out at each energy. 
Those readings along with the pieces of evidence discussed below are used to argue that we can establish the absolute charge state fractions using our relative normalization method. Furthermore, it is shown that conventional CSD measurements via cup readings are consistent with the novel method using a scintillation screen. 

\subsection{Saturation effects at $1.75$ MeV/u}
\label{sec:linearity}

At the lowest energy, $E_{lab} = 1.75$ MeV/u, a partial sweep of the charge states was completed with an attenuator inserted to reduce the total beam intensity delivered to the target by roughly a factor of three. Saturation was thought to be present online, but was not confirmed until the offline analysis. Saturation of the CMOS appears when individual pixels values have the maximum grey scale value of $255$. None of the other energies showed signs of saturation. The attenuated data did not display saturation, but images were not acquired for a significant fraction of the charge states, $q=27\text{-}29$. However, these charge states were also found to be free from saturation for the unattenated runs. As a result the two data sets could be combined and used to extract the full CSD. 

\subsection{Charge State Equilibrium}

\begin{figure}
    \centering
    \includegraphics[width=0.5\textwidth]{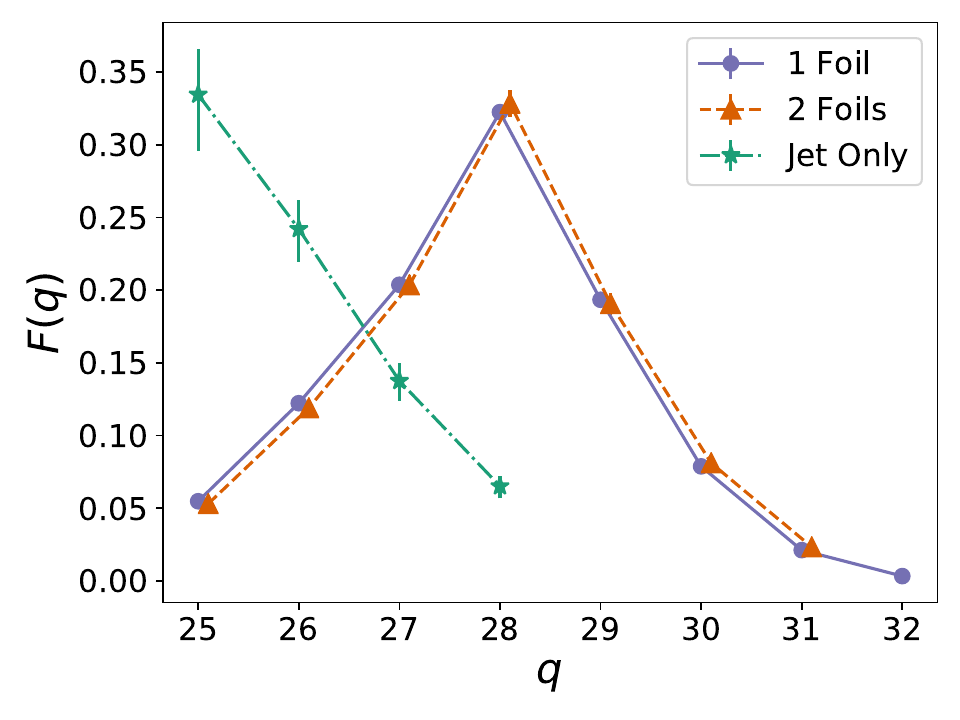}
    \caption{\label{fig:1foil_2foil_jet_comp} Measured CSDs for the post target foil (1 Foil), target foil plus post target foil (2 Foil), and just the $\sim 1.38$ MPa He jet. The 2 Foil CSD has been shifted over by $q = 0.1$ to ease the comparison. It can be seen that doubling the amount of carbon does not significantly change the CSD, indicating charge state equilibrium has been reached. The CSD for the jet has been estimated using the normalization of the other two measurements for sake of comparison, but without measuring more of the distribution the absolute scale cannot be established with any accuracy.}
    
\end{figure}

It is not clear that the $^{\text{nat}}$C foil is sufficiently thick to guarantee charge state equilibrium. Detailed study of the required thickness to achieve equilibrium would require a set of foils spanning a wide range of thicknesses. No such attempt was made for this experiment, instead an additional and identical carbon foil was installed on a $5 \times 5$ mm target frame. This frame was installed on a motorized drive located upstream of the gas jet. Inserting the second foil effectively doubled the thickness of carbon seen by the $^{88}$Sr beam. Only the scintillation screen was used for these measurements. Nearly identical CSDs were measured for both the one and two foil case as shown in Fig.~\ref{fig:1foil_2foil_jet_comp}. Using this result, it would appear that we are in equilibrium when using carbon foils of at least $42$ $\mu$g/cm$^2$.

To verify the observed distributions were due to the carbon foils, an additional test was performed at $2.75$ MeV/u by retracting all of the carbon foils, and then measuring a CSD with just the jet. A clear shift was seen towards lower charge states; however, this puts the majority of the CSD out of SECAR's rigidity limit of $0.8$ Tm. Thus, our CSD  was incomplete, and as a consequence of our relative normalization procedure, unknown. The carbon foil has the expected effect of shifting the CSD towards higher charge states \cite{sayer_1977, VOCKENHUBER_2007}.

\subsection{Effect of Helium Jet}

\begin{figure}
    \centering
    \includegraphics[width=0.5\textwidth]{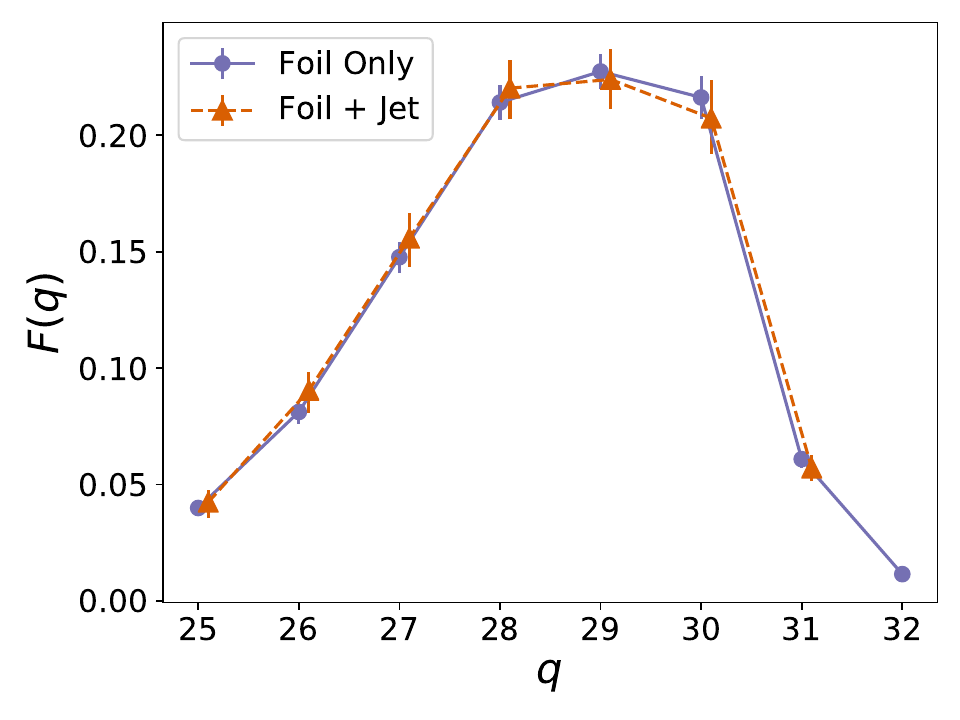}
    \caption{\label{fig:jet_on_off} Measured CSDs using the scintillation screen after the post target foil with and without the $1.38$ MPa He jet at $2.25$ MeV/$u$. The foil plus jet CSD has been shifted over by $q = 0.1$ to ease the comparison.}
    
\end{figure}

A post target stripper foil was primarily installed to remove the effects of a location-dependent charge state distribution, since non-resonant reactions can occur anywhere in the extent of the helium jet. Thus, even if charge state equilibrium is not reached, the primary concern of this experiment was whether the carbon foil completely dictated the charge state distribution. To this end, at $E = 2.25$ MeV/u a CSD was measured with both the jet on and off. The compressor discharge pressure was set to $\sim 1.38$ MPa. Using the scintillation screen, the two CSDs were found to be identical, as can be seen in Fig.~\ref{fig:jet_on_off}. The jet is known to shift the CSD down as shown in Fig.~\ref{fig:1foil_2foil_jet_comp}, so the identical distributions in Fig.~\ref{fig:jet_on_off} provide further evidence that the observed charge state distribution is due only to the carbon foil. 

\subsection{Final Results}

We present the results of our measurements in Table~\ref{tab:csd_all}. Good agreement is found between the cup and scintillation screen measurements. Due to the unknown response of the CMOS, at this time we choose to be conservative and assign an additional systematic error of $4 \%$ to the CSDs for $2.25$ and $2.75$ MeV/u. The $4 \%$ is an estimate based on the residuals between the Faraday cup and scintillation screen for the two energies. It accounts for the possibilities of overestimated uncertainties on the Faraday cup and outliers in the viewer data. For the $1.75$ MeV/u data, this same estimate resulted in a significantly higher value of $20 \%$. We take this as an indication that combining the attentuated and unattentuated data sets comes with significant additional uncertainties, possibly pointing towards issues with the linearity of the CMOS. 
%
%

\begin{figure*}
    \centering
    \includegraphics[width=\textwidth]{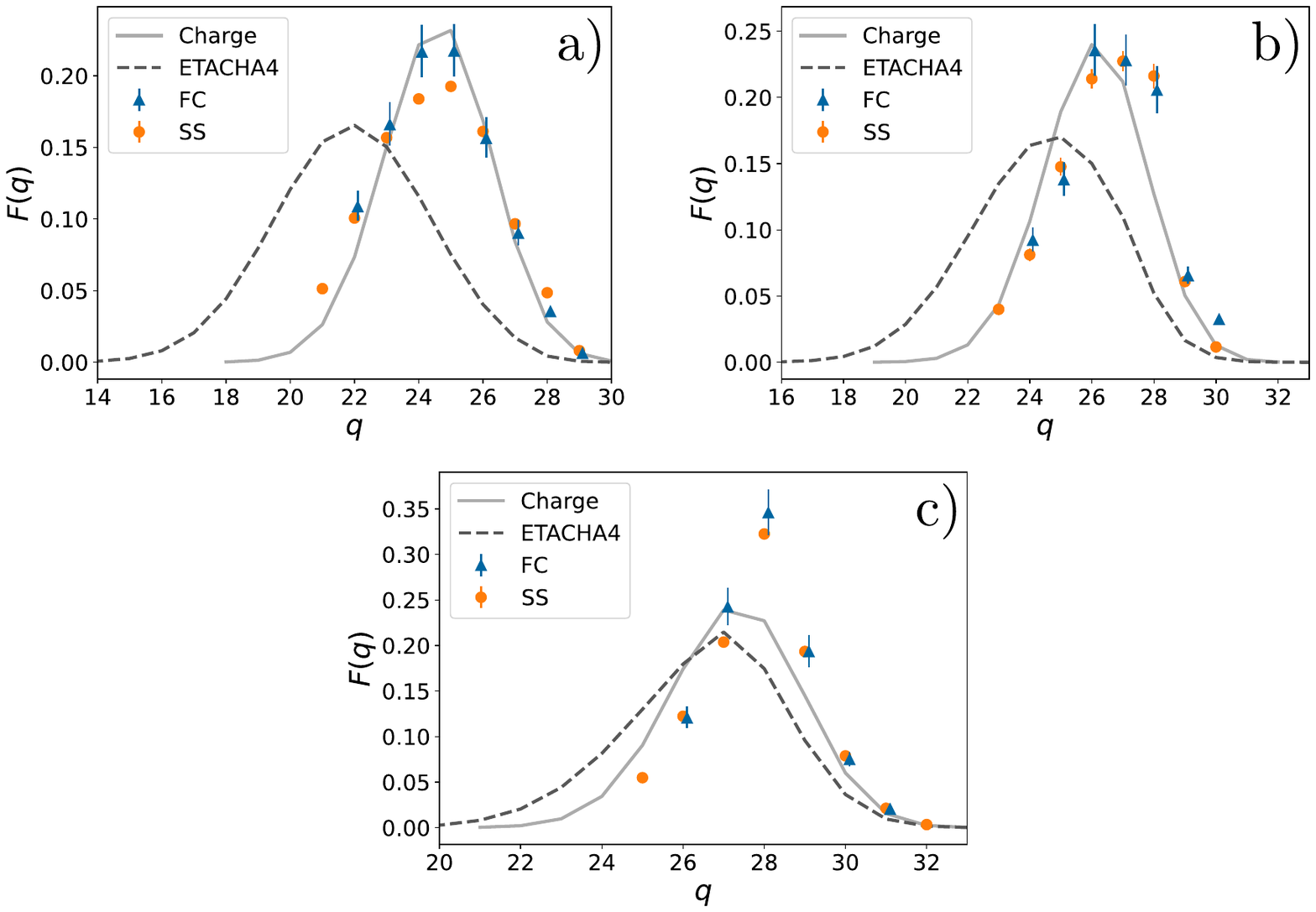}
    \caption{Comparison of measurements using Faraday cups (FC) and the scintillation screen (SS) at a) $1.75$ MeV/u, b) $2.25$ MeV/u, and c) $2.75$ MeV/u. Measurements are compared to theoretical predictions from \texttt{ETACHA4} and \texttt{Charge}. Note that the error bars are statistical only and are, in some cases, smaller than the markers.}
    \label{fig:all_comp}
\end{figure*}

Comparisons of Faraday cup and scintillation screen measurements along with theoretical predictions of the codes \texttt{ETACHA4} \cite{lamour_2015} and \texttt{CHARGE} \cite{charge} are shown in Fig.~\ref{fig:all_comp} for each beam energy. \texttt{CHARGE} appears to better describe our data, but is still insufficiently accurate, especially at the higher beam energies.

\begin{sidewaystable}[htbp]
  \centering
  \scriptsize
  \setlength\tabcolsep{5pt}
  \def\arraystretch{1.5}

  \caption{Charge state distributions as measured with scintillation screen (SS) and Faraday cups (FC). Only statistical uncertainties are reported in the table. A $4 \%$ systematic uncertainty should also be considered for the scintillation screen at $2.25$ and $2.75$ MeV/u, and a $20 \%$ systematic uncertainty at $1.75$ MeV/u.} 
  \label{tab:csd_all}
\begin{tabular}{llllllllllllll}
\hline
\hline
 Energy (MeV/u) &  Method &              $F_{21}$ &            $F_{22}$ &            $F_{23}$ &            $F_{24}$ &              $F_{25}$ &               $F_{26}$ &               $F_{27}$ &               $F_{28}$ &               $F_{29}$ &              $F_{30}$ &              $F_{31}$ &                 $F_{32}$ \\

           $1.75$ &  SS & $4.4^{+0.3}_{-0.3}$ & $8.8^{+0.5}_{-0.4}$ & $14.4^{+0.6}_{-0.6}$ & $18.7^{+0.6}_{-0.6}$ & $20.2^{+0.8}_{-0.8}$ & $17.2^{+0.8}_{-0.8}$ & $10.1^{+0.7}_{-0.7}$ & $5.1^{+0.4}_{-0.4}$ & $0.85^{+0.08}_{-0.07}$ & & & \\
             &      FC &                      &            $10.9^{+1.1}_{-1.0}$ &            $16.6^{+1.5}_{-1.4}$ &            $21.7^{+1.9}_{-1.8}$ &              $21.7^{+1.9}_{-1.8}$ &               $15.6^{+1.5}_{-1.4}$ &                 $9.0^{+0.9}_{-0.8}$ &                 $3.5^{+0.4}_{-0.3}$ &                $0.63^{+0.07}_{-0.06}$ &                      &                      &                         \\
           $2.25$ &  SS &                      &                    &   $4.0^{+0.4}_{-0.3} $ &   $8.1^{+0.5}_{-0.5}$ &    $14.8^{+0.7}_{-0.6}$ &     $21.4^{+0.8}_{-0.7}$ &     $22.7^{+0.8}_{-0.8}$ &     $21.6^{+0.9}_{-0.9}$ &      $6.1^{+0.4}_{-0.4}$ &  $1.16^{+0.09}_{-0.08}$ &                      &                         \\
             &      FC &                      &                    &                    &             $9.2^{+1.0}_{-0.9}$ &              $13.8^{+1.3}_{-1.2}$ &                  $23.5^{+2.0}_{-1.9}$ &                  $22.8^{+2.0}_{-1.9}$ &                  $20.6^{+1.8}_{-1.7}$ &                 $6.5^{+0.7}_{-0.6}$ &                $3.2^{+0.4}_{-0.3}$ &                      &                         \\
           $2.75^{\ddagger}$ &  SS &                      &                    &                    &                    &  $5.48^{+0.10}_{-0.10}$ &  $12.22^{+0.15}_{-0.15}$ &  $20.37^{+0.18}_{-0.18}$ &  $32.24^{+0.22}_{-0.21}$ &  $19.34^{+0.18}_{-0.18}$ &  $7.88^{+0.10}_{-0.10}$ &  $2.12^{+0.03}_{-0.04}$ &  $0.337^{+0.007}_{-0.007}$ \\
             &      FC &                      &                    &                    &                    &                      &               $12.0^{+1.2}_{-1.2}$ &               $24.2^{+2.1}_{-2.0}$ &               $34.6^{+2.6}_{-2.5}$ &               $19.3^{+1.8}_{-1.7}$ &                $7.5^{+0.8}_{-0.7}$ &              $2.03^{+0.23}_{-0.21}$ &                         \\
\hline
\hline

\end{tabular}
\captionsetup{labelformat=empty}

\end{sidewaystable}

\section{Conclusion}

We have developed a novel method to measure charge state distributions using a scintillation screen and used it to measure the CSD for $^{88}$Sr passing through a $42$ $\mu$g/cm$^2$ carbon foil. The method relies on the ability to simultaneously viewing multiple charge states, allowing for a relative normalization and subsequent extraction of charge state fractions. A Bayesian model was developed and applied to provide a self-consistent statistical model for this process. Further study is required to better understand and reduce possible systematic errors (estimated to be between $ 4 \%$ and $20 \%$.) related to the linearity of the scintillation screen and CMOS combination. Our method is found to be in good agreement with a traditional method using Faraday cups, although images must be carefully examined for saturation effects. However, unlike Faraday cups the scintillation screen method deals with beam intensities on the order of $10^5$ pps with ease. Our scintillation screen method is in principal similar to methods using position-sensitive parallel-plate avalanche counters \cite{KANAI_1987, GASTIS_2016}, but does not require a gas handling system or readout electronics. CSD measurements utilizing SECAR can employ complementary techniques to better constrain charge state fractions for absolute cross section measurements.

\section*{Acknowledgments}

The authors would like to thank Sam Nash and Ana Henriques for their critical support during the experiment, and to the rest of the ReA3 staff. Additional thanks to Kiana Setoodehnia for her helpful comments on the manuscript. This material is based upon work supported by the U.S. DOE, Office of Science, Office of Nuclear Physics under contract DE-AC05-00OR22725 (ORNL) and Science grants: DE-FG02-88ER40387 (OU) and DE-FG02-93ER40789 (CSM). Support was also provided from the National Science Foundation through Grants No. PHY-2011890 (UND) and PHY-1913554, PHY-2209429 (MSU). SECAR is supported by the U.S. Department of Energy, Office of Science, Office of Nuclear Physics, under Award Number DE-SC0014384 and by the National Science Foundation under grant No. PHY-1624942 with additional support from PHY 08-22648 (Joint Institute for Nuclear Astrophysics) and PHY-1430152 (JINA-CEE).

\appendix

\section{One Step Normalization Model}

A second possibility was explored for extracting charge state fractions from the measured scintillation screen intensities, as an alternative to the model presented in Sec.~\ref{sec:bay-norm-model}. The idea was to incorporate the constraint of Eq.~\ref{eq:csd_sum} directly into the model through the use of a Dirichlet prior on every $F_q$. 
\begin{equation}
    \label{eq:mult_nom}
    p(F_i) = \frac{\Gamma(\sum_{i=1}^m \alpha_i)}{\prod_{i=1}^m \Gamma(\alpha_i)} \cdot \prod_{i=1}^{m} F_i^{\alpha_i - 1},
\end{equation}
where the hyper-parameter $\alpha_i$ is called the concentration and controls how strongly the distribution is biased towards each $F_i$. Setting every $\alpha_i = 1$ produces a distribution that is uniform over the simplex of the $F_i$. 
Sampling directly from this distribution using MCMC is challenging, but the method of Ref.~\cite{betancourt_2012} allows a straight forward implementation.

Again, we normalize each image to the most intense peak. Each image has intensity varies from $0$ to $1$, implying that $\phi_k$ can take on any value from $0$ to $1$ due to the bounds of $F_q$.

The model is thus:
\begin{align}
  \label{eq:csd_model}
 & \textnormal{Priors:} \nonumber \\
 & \phi_{k} \sim \textnormal{Uniform}(0, 1) \nonumber \\
 & F_{q} \sim \textnormal{Dirichlet} (\boldsymbol{\alpha})  \nonumber \\
 & \lambda \sim \textnormal{HalfNorm} (0.1^2) \nonumber \\
 & \textnormal{Functions:}  \\
 & F_{q, k}^{exp} = \phi_k I_{q, k}  \nonumber \\
 & \textnormal{Likelihood:} \nonumber \\
 & F_{q, k}^{exp} \sim \mathcal{N}(F_{q}, \{ \lambda F_{q} \}^2)  \nonumber
 \end{align}

\begin{figure}
    \centering
    \includegraphics[width=0.5\textwidth]{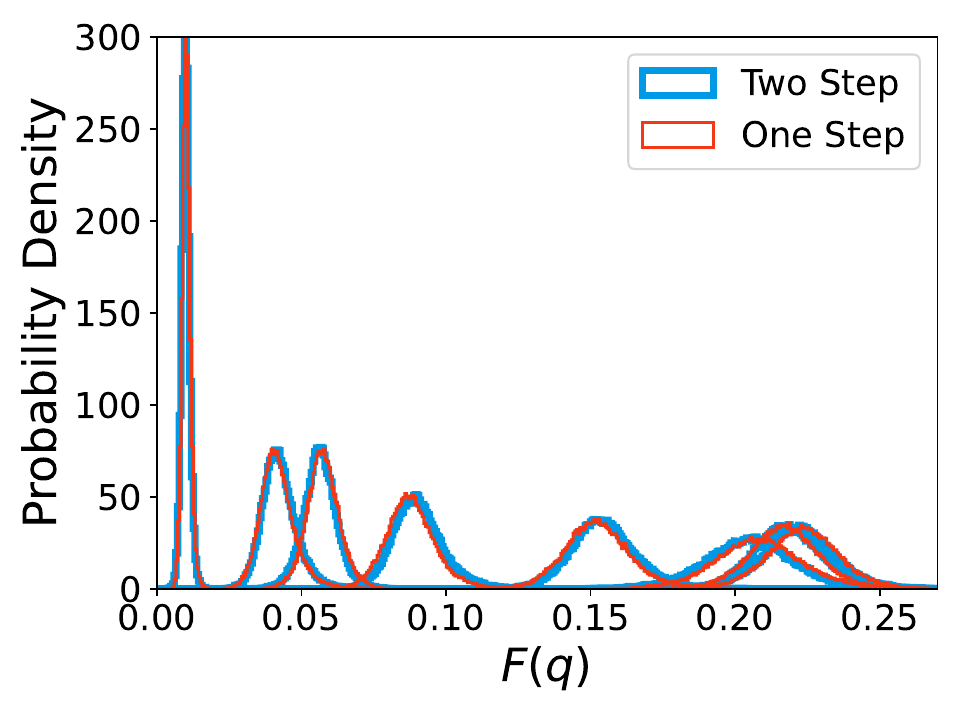}
    \caption{Comparison of one step and two step normalization models for 2.25 MeV/u $^{88}$Sr with the post-target carbon foil and the He jet on.}
    \label{fig:one_two_comp}
\end{figure}

Examining the posterior distributions for the charge state fractions in Fig.~\ref{fig:one_two_comp} shows excellent agreement between the two methods for the specific case. Due to the identical results but more complex nature of the one-step normalization process, we did not explore it further for the current work.

 \bibliographystyle{elsa-test} 
 \bibliography{refs}





\end{document}